\documentclass[twocolumn]{jpsj2}

\def\runauthor{R. Derian, A. Gendiar, and T. Nishino}

\usepackage{graphicx}
\usepackage{dcolumn}
\usepackage{bm}

\newcommand{\si}{\sigma} 

\title{Modulation of Local Magnetization
in two-dimensional Axial-Next-Nearest-Neighbor Ising model}

\author{Ren\'e Derian$^1$, Andrej Gendiar$^2$, Tomotoshi Nishino$^3$}
\inst{$^1$Institute of Physics, Slovak Academy of Sciences,
SK-845 11 Bratislava, Slovakia\\
$^2$Institute of Electrical Engineering, Slovak Academy of Sciences,
SK-841 04 Bratislava, Slovakia\\
$^3$Department of Physics, Faculty of Science, Kobe University,
Kobe 657-8501, Japan}

\recdate{\today}

\abst{
The axial next-nearest-neighbor Ising model is studied in two
dimensions at finite temperature using the density matrix
renormalization group. The model exhibits phase
transition of the second-order between the antiphase in 
low temperature and the modulated phase in high temperature.
Observing the domain wall free energy, we confirm that the
modulation period in high-temperature side is well explained 
by the free-fermion picture.
}

\kword{incommensurate phase, ANNNI model, DMRG}

\begin{document}
\maketitle

\section{introduction}

Periodically modulated structures may occur in a wide range
of physical systems. As examples of such systems,
La$_6$Ca$_8$Cu$_{24}$O$_{41}$ and Ca$_2$Y$_2$Cu$_5$O$_{10}$
are well known~\cite{Matsuda96,Fong99},
where spins of the copper atoms interact ferromagnetically
between the neighboring sites along the CuO$_2$ chains and
antiferromagnetically between the next-nearest-neighboring ones.
A phase transition of commensurate-incommensurate type was
observed in these systems. Another example is cerium antimonide
(CeSb)~\cite{Rossat77} which has a nontrivial phase diagram
and which shows modulated spin patterns with various
periodicities. In some ferroelectric materials, such as
NaNO$_3$, the modulated phases are present between the
ferroelectric low-temperature state and the paraelectric
high-temperature one~\cite{Yamada63,Massida89}.

Physical properties of magnetically modulated structures can be
described by simplified models with competing interactions.
One of the simplest examples is the
so-called axial next-nearest-neighbor Ising (ANNNI) model,
which contains ferromagnetic coupling $J_1^{~}$ between
nearest-neighbor spin pairs and antiferromagnetic one $J_2^{~}$
between next-nearest-neighbor spin pairs
in a preferred direction.~\cite{Selke88}.
Several analytical methods have been developed to study the
phase diagram of the ANNNI model in two dimensions.
For instance, the free-fermion approximation treats domain
walls running along the chain direction~\cite{VillainBak,Grynberg87}.
The 
M\"uller-Hartmann-Zittartz approach assumes existence of the
domain wall in the perpendicular direction to the axial
one~\cite{Kroemer82}. A detailed survey of earlier works on
this topic has been reviewed by Selke~\cite{Selke88}.
Recent progress can be found in
Refs. [10-12].

\begin{figure}[tb]
\includegraphics[width=\columnwidth]{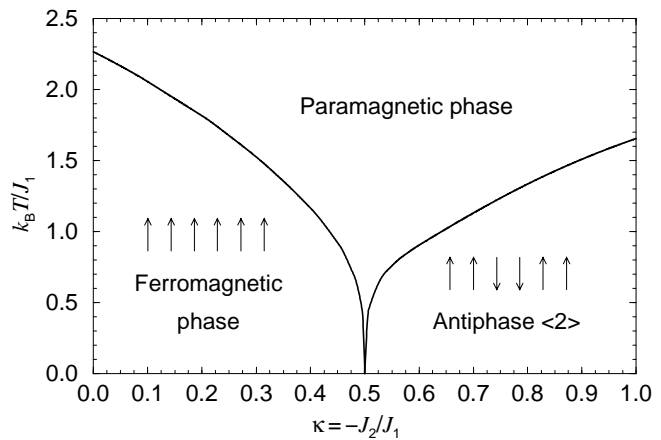}
\caption{\label{fig:1} The ordered phases of the 2D ANNNI model.}
\end{figure}

In this paper we focus on the two-dimensional (2D) 
ANNNI model, which is described by the Hamiltonian
\begin{equation}
{\mathcal H} =- J_1 \sum_{i,j} \si_{i,j}(\si_{i+1,j}
+\si_{i,j+1}) - J_2 \sum_{i,j} \si_{i,j}\si_{i+2,j}
\end{equation}
on a square lattice, where the
index $i$ specifies the position along the axial direction.
The Ising spins $\si_{i,j}=$ $\uparrow$ or $\downarrow$
interact ferromagnetically ($J_1>0$) between the nearest
neighbors and antiferromagnetically ($J_2<0$) between the
next-nearest neighbors. The ratio between the coupling 
constants $\kappa=-J_2/J_1$ is commonly used for the 
measure of the frustration. 
It is widely accepted that in the low temperature region 
the model shows a ferromagnetic structure when 
$\kappa<0.5$, and when  $\kappa$ is larger
than $0.5$, the so-called antiphase structure $\{\cdots\uparrow
\uparrow\downarrow\downarrow\uparrow\uparrow\cdots\}$
is realized.~\cite{Shirahata,VillainBak,Kroemer82,Selke88,
Grynberg87,Saqi,Murai,Sato99}
Figure~\ref{fig:1} shows the location of these ordered phases.
It has been confirmed that these ordered phases are bordered  
by the second order phase transition lines.

There is an argument about the presence of  incommensurate (IC)
phase in the highly frustrated region, which is specified by the
condition $\kappa > 0.5$.
Though a wide area of the IC phase is expected by the 
mean-field theory, the Monte Carlo (MC) simulation by 
Sato and Matsubara suggests that the region of the IC 
phase is very small.~\cite{Sato99}. 
Recently, Shirahata and Nakamura performed an extensive 
calculation by use of the non-equilibrium relaxation method~\cite{Shirahata}. 
Assuming the presence of the Berezinskii-Kosterlitz-Thouless (BKT) 
transition~\cite{Ber,KT} they estimated two critical
temperatures bordering the IC phase. What they found is that
these two transition temperatures are  almost identical. 
They speculated that successive phase
transitions may occur within an infinitesimally narrow
temperature region. Table 1 summarizes these theoretical and
numerical estimates of the phase transition temperatures at 
$\kappa=0.6$, where the $T_{\rm c}^{~}$ represents the upper border
of the antiphase, and where $T'$ is the lower border of the
paramagnetic phase. (The IC phase is present if $T'$ is larger
than $T_{\rm c}^{~}$.)
\begin{table}[b]
\label{tab:1}
\caption{Critical temperatures  at $\kappa=0.6$ known so far.}
\begin{center}
\begin{tabular}{lll}
\hline
Method used & $T_{\rm c}$ & $T'$ \\
\hline
M\"uller-Hartmann-Zittartz~\cite{Kroemer82} &
$ 1.09 $    & ---   \\
Phenomenological renorm.~\cite{Grynberg87} &
$ 1.05 $    &  $ 1.35 $ \\
Saqi and McKenzie~\cite{Saqi} &
$ 1.05 $    &  $ 1.40 $ \\
Cluster variation method~\cite{Murai} &
$ 0.91 $    &  $ 1.64 $ \\
Cluster heat bath method~\cite{Sato99} &
$ 0.91 $    &  $ 1.16 $ \\
Free-fermion approximation~\cite{VillainBak} &
$ 0.907 $   &  $ 1.20 $ \\
Non-equilibrium Relaxation~\cite{Shirahata} &
$ 0.89(2) $ &  $ 0.895(25) $ \\
DMRG (this work) &
$ 0.907 $       & ---   \\
\hline
\end{tabular}
\end{center}
\end{table}

The aim of our study is to obtain the precise modulation 
period of the local magnetization and its decay factor 
in the parameter
region where the presence of IC phase has been discussed.
For this purpose we employ the density matrix
renormalization group (DMRG)~\cite{White92,Nishino95,Schollwoeck}
method, and carry out a scaling analysis on the domain-wall
free energy. As shown in the following, we confirm that the
modulation period is well explained 
by the free-fermion picture.

\section{Application of DMRG}

We consider the 2D ANNNI model on the square lattice
of the size $L\times\infty$. The
transfer matrix of this system ${\cal T}_L^{~}[\si^\prime\vert\si]$
connects two adjacent spin rows $[\si^\prime]\equiv\{\si_{1,j}^{~},
\si_{2,j}^{~},\dots,\si_{L,j}^{~}\}$ and $[\si]\equiv\{\si_{1,j-1}^{~}$,
$\si_{2,j-1}^{~},\dots,\si_{L,j-1}^{~}\}$, where 
index $i$ runs from 1 to $L$ toward the axial direction.
For simplicity, we drop out the
indices $j$ and $j-1$  from the Ising spin variables in the following,
and write them as $[\si^\prime]\equiv\{\si'_{1},
\si'_{2},\dots,\si'_{L}\}$ and $[\si]\equiv\{\si_{1}^{~}$,
$\si_{2}^{~},\dots,\si_{L}^{~}\}$. Without
loss of generality, the transfer matrix can be written as the
product of the overlapped local weights
\begin{equation}
{\cal T}_L^{~}[\si^\prime\vert\si] = \prod_{i=1}^{L-2}
W(\si^\prime_{i}\si^\prime_{i+1}\si^\prime_{i+2}\vert\si_{i}^{~}
\si_{i+1}^{~}\si_{i+2}^{~})\, ,
\end{equation}
where $W(\si^\prime_{i}\si^\prime_{i+1}\si^\prime_{i+2}\vert
\si_{i}^{~}\si_{i+1}^{~}\si_{i+2}^{~})$ is the local Boltzmann weight
associated with the Hamiltonian ${\mathcal H}$ in Eq.~(1).~\cite{ag1,Barber}.

The DMRG is employed to solve the eigenvalue problem
\begin{equation}
\sum_{[\si]} {\cal T}_L^{~}[\si^\prime\vert\si]\ \Psi_L^{~}[\si] =
\lambda_L^{~}( T ) \, \Psi_L^{~}[\si^\prime]
\end{equation}
with $\lambda_L^{~}( T )$ is the largest
eigenvalue of the transfer matrix and $\Psi_L^{~}[\si]$ the
corresponding eigenvector. We employ two different boundary
conditions: the parallel ones ($\si_{1}^{~}=\si'_{1}=\uparrow$ and
$\si_{L}^{~}=\si'_{L}=\uparrow$) and the antiparallel ones
($\si_{1}^{~}=\si'_{1}=\uparrow$ and $\si_{L}^{~}=\si'_{L}=
\downarrow$), respectively, for which we calculate the largest eigenvalues
$\lambda_L^{\uparrow\uparrow}( T )$ and $\lambda_L^{\uparrow
\downarrow}( T )$. For the visualization of the
spin modulation, we calculate the local magnetization
\begin{equation}
\left\langle \si_i \right\rangle=
\frac{\sum_{[\si]}\Psi_L^{~}[\si]\si_i\Psi_L^{~}[\si]}
     {\sum_{[\si]}\Psi_L^{~}[\si]     \Psi_L^{~}[\si]}
\end{equation}
as a function of position $i$ during the last sweep 
in the zipping process of the finite-system 
DMRG~\cite{White92}. We keep at most $m=70$ block-spin
states and vary the lattice size from $L = 38$ to $L = 118$.
Note that under these conditions the density matrix
truncation error~\cite{White92,Nishino95,Schollwoeck}
is kept within $\varepsilon\leq10^{-13}$.

\begin{figure}[tb]
\includegraphics[width=\columnwidth]{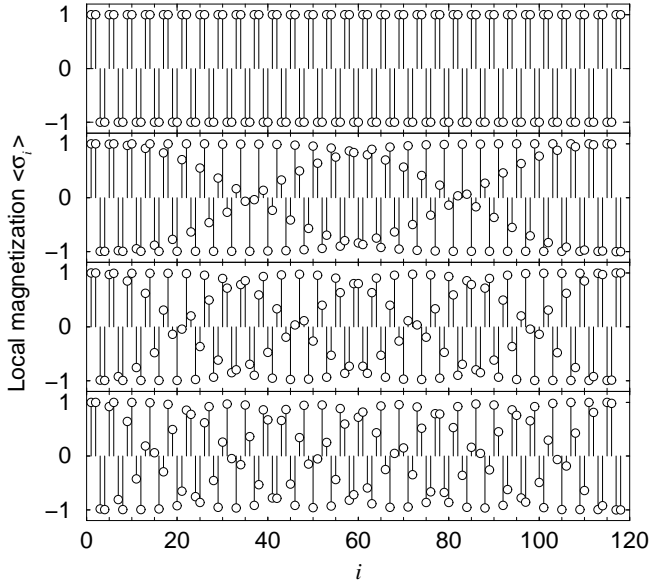}
\caption{The local magnetization $\langle\si_i\rangle$
 calculated for $L =118$
($i = 1,2,\dots,118$) with parallel and antiparallel
boundary conditions. }
\label{fig:2}
\end{figure}

 We use dimensionless units  $k_{\rm B}$\,=\,$J_1$\,=\,$1$
throughout this article. 
We focus on analysis of the model at $\kappa=0.6$, where the 
competing interaction plays an important role on the spin modulation.
Figure~\ref{fig:2} shows the local magnetization $\langle\si_i
\rangle$ at $\kappa=0.6$ under and over a transition
temperature $T_{\rm c}\approx0.91$ which we will
determine more precisely. The complete antiphase structure
$\{\uparrow\uparrow\downarrow\downarrow\}$ is observed at $T=0.88$
if the parallel boundary conditions are imposed (the uppermost)
and a twisted pattern created by a running domain wall is observed
for the antiparallel conditions (the second from top). The
remaining two panels display $\langle\si_i\rangle$ at $T=0.93$,
where a modulated structure is present for the parallel conditions (the
third panel) and the antiparallel ones (the fourth). Note that the
modulation period depends on the applied boundary conditions.

\section{Modulation Period}

\begin{figure}[tb]
\includegraphics[width=\columnwidth]{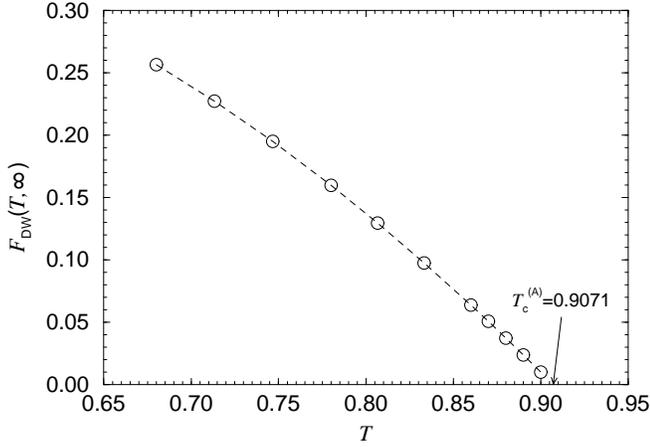}
\caption{The domain-wall energy
${\cal F}_{\rm DW}(T,\infty)$ at $\kappa= 0.6$.}
\label{fig:3}
\end{figure}

For the purpose of characterizing the spin modulation, we
introduce the ``domain-wall free energy''~\cite{dwefe}
\begin{equation}
{\cal F}_{\rm DW}(T,L) = (-1)^{n(L)} k_{\rm B}T\ln
\frac{\lambda_L^{\uparrow\downarrow}(T)}{\lambda_L^{\uparrow
\uparrow}(T)} \, ,
\end{equation}
where $n(L) = [L/2+2]+L+1$ represents the 4-site periodicity in the
antiphase. The ${\cal F}_{\rm DW}(T,L)$ 
represents the sensitivity of the free energy per
lattice row to the boundary conditions. In the antiphase
region, ${\cal F}_{\rm DW}(T,L)$ exhibits the $L$
dependence
\begin{equation}
{\cal F}_{\rm DW}(T,L) \sim {\cal F}_{\rm DW}(T,\infty)
+ c(T) L^{-2},
\end{equation}
where $c(T)$ is a parameter, which is related to the 
`mass' of the moving domain wall, and where
${\cal F}_{\rm DW}(T,\infty)$ represent the
 `stationary' domain-wall energy. 
Figure~\ref{fig:3} shows the
dependence of ${\cal F}_{\rm DW}(T,\infty)$ with
respect to $T$. The domain-wall free energy vanishes
at critical temperature $T_{\rm c}^{\rm(A)}=0.907$;
the superscript (A) stands for the fact that we have estimated 
the temperature inside the antiphase region.

\begin{figure}[tb]
\includegraphics[width=\columnwidth,clip]{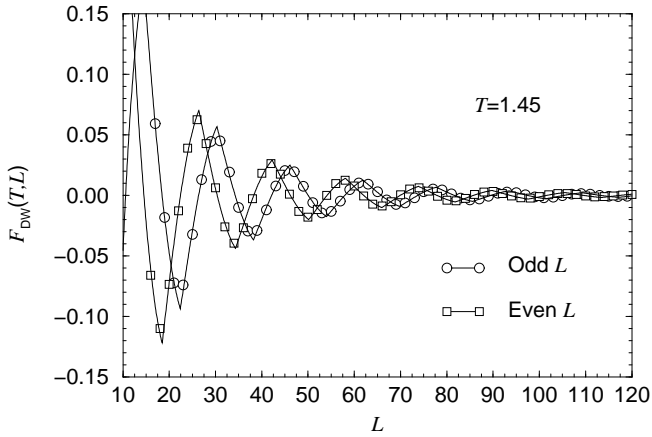}
\caption{The domain-wall free energy with respect to size $L$ at
$T=1.45$ and $\kappa=0.6$ (the symbols). The
saw-like fitting functions are given by Eq.~\eqref{JagFnc}.}
\label{fig:4}
\end{figure}

\begin{figure}[tb]
\includegraphics[width=\columnwidth,clip]{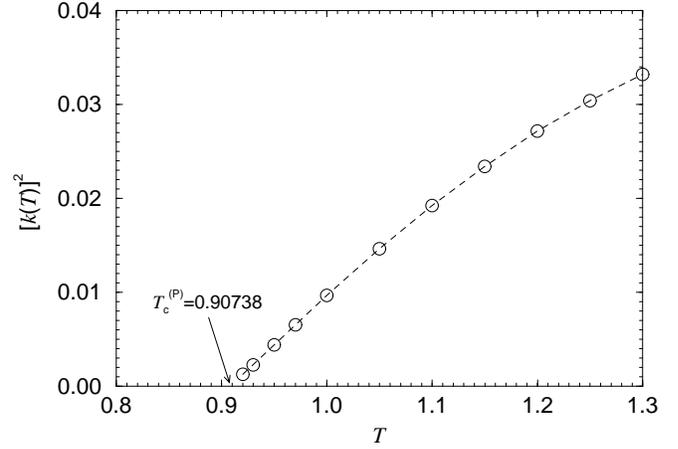}
\caption{Behavior of the wave number $[ k(T) ]^2
={[q(T) - \pi/2]}^2$.}
\label{fig:5}
\end{figure}

A detailed analysis of the ${\cal F}_{\rm DW}(T,L)$ is required
in the higher temperature region ($T>T_{\rm c}$). The typical
$L$-dependence of ${\cal F}_{\rm DW}(T,L)$ is depicted in
Fig.~\ref{fig:4}. The saw-like structure in  ${\cal F}_{\rm DW}(T,L)$ is 
naturally explained from the fact that the system prefers to have 
`a natural' wave number $q(T)$ for the spin modulation 
if the system size is infinitely large ($L\to\infty$). 
When $L$ is finite, the boundary conditions force
the system to have a modified wave number $q^\prime(T)$, which
is quantized as
\begin{eqnarray}
\nonumber
q^\prime_{\uparrow\uparrow}(T)&=&2\pi m/(L-\ell)\, ,\\
q^\prime_{\uparrow\downarrow}(T)&=&2\pi(m+1)/(L-\ell) \, ,
\label{qntEq}
\end{eqnarray}
respectively, for the parallel and the antiparallel conditions,
where $m$ is an appropriate integer and $\ell$ is an offset. 
As a
consequence of the `forced' shift in the wave number, a small
increase of the free energy per site occurs and is proportional to
${[q(T)-q^\prime_{\uparrow\uparrow}(T)]}^2 $ and
${[q(T)-q^\prime_{\uparrow\downarrow}(T)]}^2 $ if
higher-order corrections are omitted. Paying attention to the
quantization condition in Eq.~\eqref{qntEq} and subtracting
${\cal F}_{\uparrow\uparrow}( T, L ) = -k_{\rm B}T
\ln \lambda_L^{\uparrow\uparrow}(T)$ from
${\cal F}_{\uparrow\downarrow}( T, L ) = -k_{\rm B}T
\ln \lambda_L^{\uparrow\downarrow}(T)$, we obtain the saw-like
dependence in ${\cal F}_{\rm DW}(T,L)$ with respect to $L$ shown
in Fig.~\ref{fig:4}. For the quantitative determination of
$q(T)$, we employ a fitting function of the form
\begin{equation}
\label{JagFnc} {\cal F}_{\rm
DW}(T,L) = \frac{a e^{-d L}}{L}
\bigl\{\left|\cos(kL+\varphi)\right|-
  \left|\sin(kL+\varphi)\right|\bigr\}
\end{equation}
that contains four temperature dependent parameters: $a$ is an
amplitude, $\varphi$ is the phase offset which is related to $\ell$ in
Eq.~\eqref{qntEq}, $d$ is a dumping, and 
\begin{equation}
k\equiv k(T)=q(T)
-\pi/2
\end{equation}
represents a change of the wave number $q(T)$ from the
antiphase wave number $\pi/2$~\cite{elsewhere}. 
(Precisely speaking, as shown in Fig.~4, there is a kind of even-odd
oscillation, and we have to shift $\varphi$ in Eq.~(8) by $\pi/4$ when 
$L$ is odd.)
In Fig.~\ref{fig:5}, we plot $\bigl[k(T) \bigr]^2$ as a function of
temperature $T$. Performing the extrapolation for $k(T)$,
we obtain the critical temperature $T_{\rm c}^{\rm (P)}=0.907$,
which is determined in the  paramagnetic region. This result is in accordance
with the previously obtained $T_{\rm c}^{\rm (A)}$. It should be
noted that the  linearity of $\bigl[k(T) \bigr]^2$
with respect to $T$ is in accordance with the free-fermionic picture
on the spontaneously created domain walls.~\cite{VillainBak,Grynberg87}.

Let us observe the temperature dependence of the decay 
factor $d(T)$. Figure 6 shows $d(T)$ in logarithmic scale.
Since our survey is limited to $L \le 118$, the estimated $d(T)$ 
for each temperature contains relatively large fitting error, which is
visible as a fluctuation of the plotted data. Among the trial 
functions we have considered, the one
\begin{equation}
d(T) = \alpha \, \exp[ -\beta ( T - T_{\rm c}^{~} )^{-1}_{~} ]
\end{equation}
shows the best fit when $\alpha = 1.32(30)$, 
$\beta = 2.17(32)$, and $T_{\rm c}^{~} = 0.907(50)$. Though
we don't have any clear picture about the reason of the 
temperature dependence, the obtained $T_{\rm c}^{~}$ is
in accordance with $T_{\rm c}^{(A)}$ and $T_{\rm c}^{(P)}$.

\begin{figure}[tb]
\includegraphics[width=\columnwidth,clip]{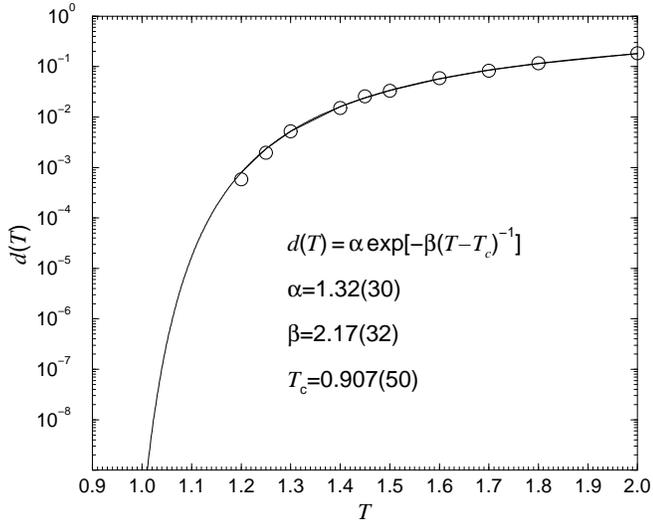}
\caption{Temperature dependence of the dumping factor.}
\label{fig:6}
\end{figure}

\begin{figure}[tb]
\includegraphics[width=\columnwidth,clip]{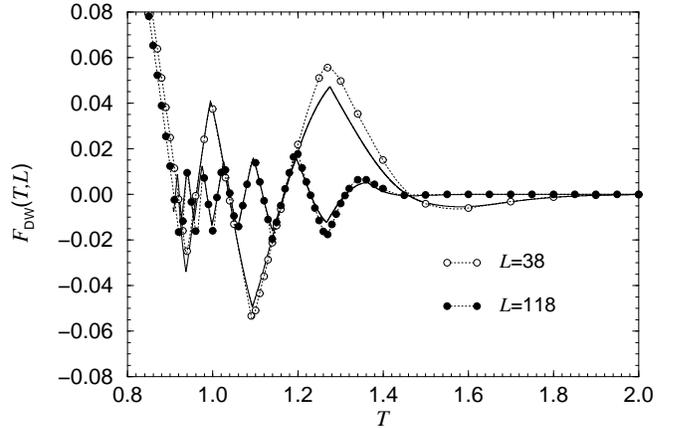}
\caption{Dependence of ${\cal F}_{\rm DW}(T,L)$ on temperature
$T$ at $\kappa=0.6$. The symbols with dotted lines
represent the calculated ${\cal F}_{\rm DW}(T,L)$ and the 
full curves are drawn by  Eq.~\eqref{JagFnc}.}
\label{fig:7}
\end{figure}

As an complementary approach, we calculate ${\cal F}_{\rm DW}
(T,L)$ as a function of $T$ at $\kappa=0.6$ (see
Fig.~\ref{fig:7}.) For simplicity we choose even $L$ in this
calculation. Again, we obtain the saw-like dependence
of ${\cal F}_{\rm DW}(T,L)$ which rapidly decays in higher
temperature region. It is expected that the 
 fitting function in Eq.~\eqref{JagFnc} with the previously
determined parameters $a$, $k$, $\varphi$, and $d$ explains
the plotted data. The curve shown in Fig.~\ref{fig:7} is thus
drawn, and a good
agreement is found between the plotted data and
the fitting by Eq.~\eqref{JagFnc}.

\begin{figure}[tb]
\includegraphics[width=\columnwidth,clip]{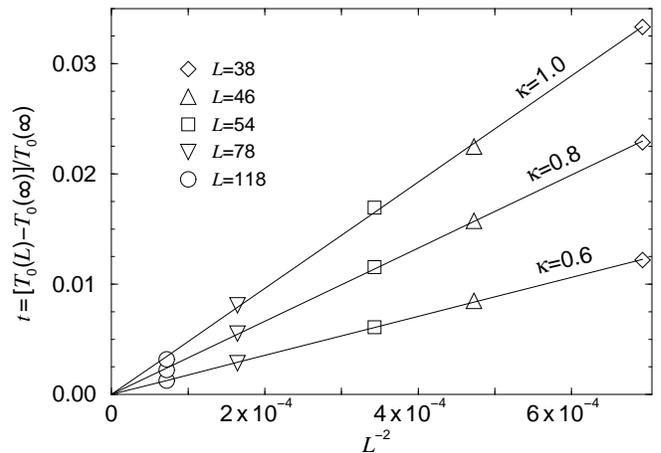}
\caption{The relative temperature $t$ with respect to $L^{-2}$
for $\kappa=0.6$, $\kappa=0.8$, and $\kappa=1.0$.}
\label{fig:8}
\end{figure}

Now, let us observe
the `first zero-crossing temperature' of ${\cal F}_{\rm DW}(T,L)$ 
from the low-temperature side, when $L$ is fixed.
Speaking phenomenologically, the
wave number in the thermodynamic limit $q(T)$ 
starts to deviate from $\pi/2$ at the
critical temperature $T_{\rm c}$. 
Therefore for the finite size system the domain wall energy ${\cal F}_{\rm DW} (T,L)$
becomes zero when
\begin{equation}
[ q(T) - \pi/2 ]L = k(T)L = \pi/2
\end{equation}
is satisfied. Thus
the zero-crossing temperature, say $T = T_0^{~}(L)$, is slightly larger than
$T_{\rm c}^{~}$. From the analysis we have performed, we 
already know that $k(T)$ is proportional to 
$\sqrt{T-T_{\rm c}}$. Combined with Eq.(11), we get the 
relation 
\begin{equation}
\sqrt{T-T_0^{~}(L)} \propto
L^{-1} \, .
\end{equation} In order to confirm this, we plot
the relative temperature
\begin{equation}
t = \frac{T_0^{~}(L)-T_0^{~}(\infty)}{T_0^{~}(\infty)}
\end{equation}
with respect to $L^{-2}$ in Fig.~\ref{fig:8}, 
where $T_0^{~}(\infty)$ for each $\kappa$ is appropriately chosen so
that the best linearity in the plotted data is realized.
As a result, we obtain $T_0^{~}(\infty)=0.907$ at $\kappa=0.6$,
which is in accordance with previously obtained transition temperatures.

For other values of $\kappa$ we obtain $T_{0}^{~}=1.335$
at $\kappa=0.8$, and $T_{0}^{~}=1.654$ at $\kappa=1.0$. The 
transition temperatures thus obtained
are used for drawing the phase diagram shown in Fig.~1.
It should be noted that the
divergence of the correlation length at the transition temperature
leads to the relation $t\propto L^{-1/\nu}$~\cite{Barber}, and from
the relation we obtain a critical exponent $\nu=0.5$.
This value is in accordance with the free-fermion 
picture~\cite{VillainBak,Grynberg91}.

\section{Conclusions and Discussions}

In conclusion, we have applied DMRG to the 2D ANNNI model. 
Observing the spin modulation period, we confirm that the free-fermion
picture well describes the phase transition from the antiphase to the
modulated state. 

Since the correlation length is far longer than the size of systems 
in our study when the temperature is slightly higher than $T_{\rm c}$, 
we cannot directly judge whether there is a `stable' IC phase 
or the modulated states are actually decaying in the long distance limit. 
At least we can say that we observe no conspicuous singularity 
in the modulation period above $T_{\rm c}$ as shown in Fig.~5.
Comparing the fact with the possible temperature region for the
IC phase reported by  Shirahata and Nakamura~\cite{Shirahata},
we conjecture that there is no IC phase in 2D ANNNI model, which
is one of the possibility pointed by Shirahata and Nakamura.

This work is supported by the VEGA grants No.
2/6071/26 and No. 2/3118/23, and the Grant-in-Aid for Scientific Research from
Ministry of Education, Science, Sports and Culture (No.~17540327).

\end{document}